\documentclass[aps,prl,reprint,superscriptaddress,amsmath,amssymb,floatfix]{revtex4-2}

\usepackage{graphicx}
\usepackage{bm}
\usepackage{siunitx}
\usepackage[colorlinks=true,linkcolor=blue,citecolor=blue,urlcolor=blue]{hyperref}

\graphicspath{{figures/}}

\newcommand{\aL}{a_\mathrm{L}}
\newcommand{\dL}{\delta_\mathrm{L}}
\newcommand{\ls}{\ell_\mathrm{s}}
\newcommand{\Es}{E_\mathrm{s}}
\newcommand{\Ef}{E_\mathrm{f}}
\newcommand{\ve}{v_\mathrm{e}}
\newcommand{\tE}{t_\mathrm{E}}
\newcommand{\Psix}{P_{60}}
\newcommand{\nhat}{\hat{\bm{n}}}

\begin{document}

\title{Drying kinematics selects open honeycomb fold networks on solvent-extracted viscoelastic thin films and gels}

\author{Fan Zhao}
\thanks{These authors contributed equally to the work.}
\affiliation{Department of Chemical Engineering, Waterloo Institute for Nanotechnology, University of Waterloo, Waterloo, Ontario N2L 3G1, Canada}

\author{Surjyasish Mitra}
\thanks{These authors contributed equally to the work.}
\affiliation{Department of Mechanical and Mechatronics Engineering, Waterloo Institute for Nanotechnology, University of Waterloo, Waterloo, Ontario N2L 3G1, Canada}

\author{Sunandan Tandon}
\affiliation{Department of Physics and Astronomy, University of Waterloo, Waterloo, Ontario N2L 3G1, Canada}

\author{Zoya Leonenko}
\affiliation{Department of Physics and Astronomy, University of Waterloo, Waterloo, Ontario N2L 3G1, Canada}
\affiliation{Department of Biology, University of Waterloo, Waterloo, Ontario N2L 3G1, Canada}

\author{Boxin Zhao}
\email[Contact author: ]{zhaob@uwaterloo.ca}
\affiliation{Department of Chemical Engineering, Waterloo Institute for Nanotechnology, University of Waterloo, Waterloo, Ontario N2L 3G1, Canada}

\author{Sushanta K. Mitra}
\email[Contact author: ]{skmitra@uwaterloo.ca}
\affiliation{Department of Mechanical and Mechatronics Engineering, Waterloo Institute for Nanotechnology, University of Waterloo, Waterloo, Ontario N2L 3G1, Canada}
\affiliation{International School of Engineering (ISE), Faculty of Engineering, Chulalongkorn University, Bangkok 10330, Thailand}

\begin{abstract}
Solvent extraction and drying of substrate-supported viscoelastic films and gels produce quasiperiodic networks of three-armed surface folds. In this Letter, we examine the physical mechanisms governing their formation, length scales, and geometric organization. We observed that swelling-induced compression in the pinned film drives localized folding of a stiffening, diffusively thickening surface skin over a drained underlayer. Combining experiments and theory, we reveal that both the pattern wavelength $\lambda$ and fold-arm length $\aL$ scale with the dried-film thickness as $\dL^{5/8}$, with remarkably weak dependence on elastic modulus. The folds form $120^\circ$ junctions on a triangular lattice, defining an open honeycomb network whose pitch satisfies $p=\sqrt{3}(\aL+g)$, where $g$ is the junction gap. Notably, the folds accommodate less than 0.1\% of the stored compressive strain, implicating elastocapillary arrest, rather than strain relaxation, in limiting their amplitude. These findings establish distinct geometric, kinetic, and interfacial constraints governing pattern selection in drying soft solids.
\end{abstract}

\maketitle

Soft solids break the symmetry of their free surface under compression. Wrinkles, creases, folds, and ridges have been mapped in elastomers, gels, and living tissue. The thresholds and wavelengths that separate these states are well understood~\cite{Bowden1998,Cerda2003,Li2012,Hong2009,Hohlfeld2011,Cai2012,Trujillo2008,Tanaka1987,Chen2004,Dervaux2012}. However, most of the existing literature concerns planforms such as stripes, labyrinths, or arrays of dimples. Far less is known about the networks a soft surface can form. Such a network consists of singular lines embedded in the surface that meet at vertices. In nature, the canonical networks of this kind are fracture networks. Columnar joints in cooling lava, as at the Giant's Causeway, and desiccation cracks in drying mud tile the plane with polygonal, often near-hexagonal cells, and repeated drying cycles drive their junctions from $90^\circ$ T-junctions toward $120^\circ$ Y-junctions~\cite{Goehring2010}. The column width is selected by the kinematics of the advancing cooling or drying front: it scales inversely with front speed at a nearly constant P\'eclet number~\cite{Goehring2009}. However, those networks form by material separation. Whether a continuous soft surface can build an analogous network by folding alone, and which underlying mechanisms then select its topology, mesh size, and local structure, remains unestablished.

\begin{figure*}[t]
\includegraphics[width=\textwidth]{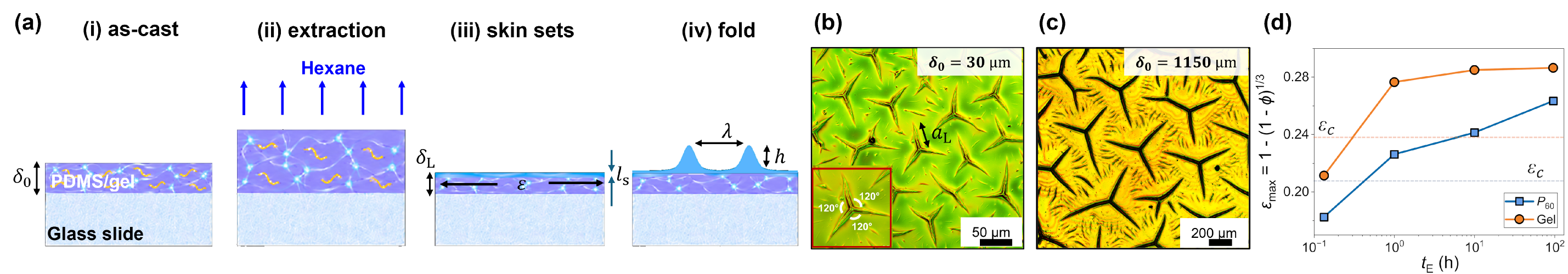}
\caption{(a) Mechanism of pattern formation on soft PDMS and gel films. (i) The as-cast film of thickness $\delta_0$ containing a finite mass fraction of uncrosslinked chains. (ii) Immersion in hexane swells the network and leaches out the mobile phase. (iii) On drying, a solvent-depleted skin of thickness $\ls$ stiffens and stops flowing while the layer is still swollen. $\dL$ is the dried film thickness. This locks in the compressive strain $\varepsilon$ that the pinned footprint imposes on any swollen layer. (iv) The skin folds with wavelength $\lambda$ and amplitude $h$. (b), (c) Reflection interference contrast microscopy images of PDMS ($\Psix$) films after \SI{96}{\hour} of hexane extraction and air drying, for initial thicknesses $\delta_0=\SI{30}{\micro\metre}$ and \SI{1150}{\micro\metre}. The arms almost meet in (b), and their length is represented as $\aL$. They have closed into an almost complete honeycomb in (c). The inset of (b) shows a single pattern with an inter-arm angle of $120^\circ$. (d) Upper bound on the released strain, Eq.~(\ref{eq:strain}) evaluated with the measured total mass loss, versus extraction time $\tE$ for the PDMS (squares) and the silicone gel (circles). The gel crosses at a critical strain $\varepsilon_\mathrm{c}\simeq0.21$ (orange solid line), and PDMS at $\varepsilon_\mathrm{c}\simeq0.23$ (blue solid line). Image (b) is adapted with permission from Ref.~\cite{Zhao2025}.}
\label{fig1}
\end{figure*}

We recently described triradial patterns, a type of such network, formed upon solvent extraction and drying of surface-attached viscoelastic and gel films~\cite{Zhao2025}. In their as-cast state, they contained a large mass fraction ($\varphi_\mathrm{f}$) of uncrosslinked, free-liquid chains. Immersion in a solvent like $n$-hexane swells the network and leaches out the mobile phase. Following air drying, the surface exhibits a quasiperiodic array of three-armed triradial folds/ridges with an interarm angle close to $120^\circ$ [Figs.~\ref{fig1}(a)--\ref{fig1}(c)]. The morphology is reproducible, surviving on glass, silicon, and cured elastomer supports [Fig.~\ref{figS1}, Supplemental Material]. Hence, it is not a delamination pattern. Further, the morphology is stable for months, and an incision does not disturb neighboring motifs, eliminating the role of residual compression in holding patterns together [Figs.~\ref{figS2}--\ref{figS3}, Supplemental Material]. However, questions remain. What fixes the topology, i.e., why are the vertices trivalent with arms at $120^\circ$? What fixes the mesh size: the arm length, the vertex spacing, and the gap at which neighbouring arms stop? And what limits the local structure, namely the ridge height? In this Letter, we answer these questions by experiments and scaling analysis. Our central structural finding is that a triradial motif is not an isolated object. It is a \textit{vertex of a honeycomb fold network} whose nodes sit on a triangular lattice.

We performed experiments with commercially available PDMS (Sylgard 184, base:\ crosslinker, 60:1, Young's modulus $E=\SI{9.6}{\kilo\pascal}$~\cite{Zhao2025}, henceforth referred to as $\Psix$) and Silicone gel (Gelest, DMS-V31, Young's modulus $E=\SI{5.6}{\kilo\pascal}$~\cite{Zhao2025}) coated on microscope glass slides, swollen in hexane and dried in air (Notes S1 and S2, Supplemental Material). Patterns were imaged by dual-wavelength reflection interference contrast microscopy (DW-RICM) and AFM, from which we extracted the pattern wavelength $\lambda$, the fold arm length $\aL$, and the network pitch $p$ and junction gap $g$ [Fig.~\ref{figS4}, Supplemental Material]. We combine our experimental findings with the kinematics of solvent extraction and a Winkler estimate of the pattern wavelength.

\textbf{\textit{Where the compression comes from.}} The network is crosslinked in the presence of mobile chains, so the as-cast film defines its stress-free reference state (see Fig.~\ref{fig1}a). Because adhesion pins the film to the substrate, however, subsequent changes in liquid content cannot be accommodated by free lateral expansion or contraction (see Fig.~\ref{fig1}a). Instead, they occur primarily through the film thickness. From thickness and gravimetric measurements, we observe that the substrate $\Psix$ experiences a mass loss of 60\% following solvent extraction and drying, yielding $\varphi_\mathrm{f}=0.60$, and contracts to a final thickness of $\dL=\SI{12.9}{\micro\metre}$. Thus, the measured ratio $\dL/\delta_0=12.9/30=0.43$ agrees with $1-\varphi_\mathrm{f}=0.40$, consistent with a nearly fixed lateral footprint~\cite{Zhao2025}. This constraint has an important consequence: when the network absorbs hexane and swells beyond its as-cast liquid content, it prefers to expand laterally but cannot and therefore develops a constraint-induced in-plane compressive strain (see Fig.~\ref{fig1}a). Conversely, once drying removes enough liquid for the network to fall below its as-cast reference content, the preferred lateral size becomes smaller than the pinned footprint, and the film tends toward tension. Thus, the compression responsible for the instability is transient: it exists while the constrained film contains more liquid than it did when crosslinked, even though little residual compression remains in the final dried state, consistent with the observation that an incision does not perturb neighboring motifs.

Drying therefore does not create compression; rather, it converts an already existing compressive strain into an elastic instability. While the film remains strongly solvent-rich, the compressive strain is largely unable to generate persistent stress. Its relaxation time $\tau=\eta/G$ is short compared with the drying time $t_\mathrm{d}$. Hence, the characteristic Deborah number is $\mathrm{De}=\tau/t_\mathrm{d}<1$, and stresses can relax by molecular rearrangement and flow~\cite{Huang2002}. At the same time, the large elastocapillary ratio $\gamma/E\delta$ suppresses sharp surface deformation~\cite{Chen2012}, where the characteristic length scale $\delta$ scales with $\delta_0$ or $\dL$. As solvent leaves the free surface, however, the local polymer concentration rises, and the surface layer stiffens dramatically, with $E$ increasing by roughly three orders of magnitude toward the megapascal modulus of dry PDMS. Consequently, $\gamma/E\delta$ decreases while the relaxation time grows. The surface eventually becomes mechanically arrested while the underlying film still contains excess liquid and therefore still carries the swelling-induced compressive mismatch. This locking-in of a compressed surface layer, not the drying itself, makes the stored compressive strain available to drive folding. Consequently, the strain available for release is set by the liquid it has yet to lose and can be expressed as,
\begin{equation}
\varepsilon = 1-(1-\Delta\varphi)^{1/3}.
\label{eq:strain}
\end{equation}

The amount of compression available when this arrest occurs depends on how much liquid remains to be lost after the surface becomes immobilized. Because the liquid content at the instant of lock-in is not directly measurable, Eq.~(\ref{eq:strain}) provides an upper bound based on the total mass subsequently removed during extraction and drying. Evaluating this bound with the measured mass losses for different extraction times ($\tE$) gives $\varepsilon=0.18$, 0.23, 0.24, and 0.26 for $\Psix$ at $\tE=\SI{8}{\minute}$, \SI{1}{\hour}, \SI{10}{\hour}, and \SI{96}{\hour} [Fig.~\ref{fig1}(d)]. The corresponding values for the gel are 0.21, 0.28, 0.28, and 0.29 [Fig.~\ref{fig1}(d)]. Ordered arrays of patterns appear in the gel already at 8 mins of solvent extraction, whereas the PDMS reveals well-ordered pattern arrays from approximately 10 hr onward [Fig.~\ref{figS5} and Table~\ref{tabS1}, Supplemental Material], placing the corresponding onset windows near $\varepsilon_\mathrm{c}\simeq0.21$ for the gel and 0.23--0.24 for the PDMS. Note that since the gel is softer and more mobile, it folds at a slightly lower strain. Furthermore, both these values lie well below the creasing strain of 0.35 for a homogeneous half-space~\cite{Hong2009,Cai2012}, as expected for a film with a strongly graded surface modulus~\cite{Trujillo2008}.

\textbf{\textit{What sets the mesh size.}} In dictating the mesh size, two lengths compete: the thickness $\ls$ of the stiffened surface skin and the film thickness $\dL$. Solvent transport alone cannot produce the skin, because hexane diffuses through swollen PDMS with diffusivity $D_\mathrm{f}\sim\SI{e-9}{\metre\squared\per\second}$~\cite{Hu2011,Phadnis2018}, yielding a solvent P\'eclet number $\ve\dL/D_\mathrm{f}\lesssim0.1$ and a nearly uniform solvent profile. The skin is instead limited by a much slower cooperative diffusivity $D_\mathrm{s}$ in the concentrated surface layer, and once formed, it thickens diffusively throughout drying~\cite{Okuzono2006,Talini2023}. For the present system, $D_\mathrm{s}$ is dictated by the self-diffusion of mobile PDMS and is of the order of \SI{e-11}{\metre\squared\per\second}~\cite{Appel1993}. Therefore, for a film receding at an approximately constant velocity $\ve$, which dries in $t_\mathrm{d}\sim\dL/\ve$, the skin thickness can be expressed as (see Note S3, Supplemental Material),
\begin{equation}
\ls \sim (D_\mathrm{s}t_\mathrm{d})^{1/2} \sim (D_\mathrm{s}\dL/\ve)^{1/2}.
\label{eq:skin}
\end{equation}

The skin is thin ($\ls\ll\dL$) and stiff ($\Es\gg\Ef$) relative to the underlying film of thickness $\dL$. Crucially, during drying, the film is \textit{drained}: it relieves normal load by expelling liquid rather than by flowing laterally and therefore responds as a compressible foundation of Winkler stiffness $K\sim\Ef/\dL$ rather than as an incompressible one~\cite{Huang2005,Brau2013}. For a stiff skin resting on a Winkler foundation, the bending stiffness is $B=\Es\ls^3/12$. Here, both $\Ef$ and $\Es$ are effective moduli normalized by $1-\nu^2$, where $\nu$ is Poisson's ratio. Minimizing the total energy with respect to wavenumber yields $\lambda=2\pi(B/K)^{1/4}$ for $\lambda\gg\dL$ [\citenum{Cerda2003},\,\citenum{Huang2005}--\citenum{Hetenyi1946} and Note S3, Supplemental Material]. With Eq.~(\ref{eq:skin}),
\begin{equation}
\lambda \simeq 2\pi\left(\frac{\Es\ls^3\dL}{12\Ef}\right)^{1/4} \sim \left(\frac{\Es}{\Ef}\right)^{1/4}\left(\frac{D}{\ve}\right)^{3/8}\dL^{5/8}.
\label{eq:lambda}
\end{equation}

\begin{figure}[tb]
\includegraphics[width=0.5\textwidth]{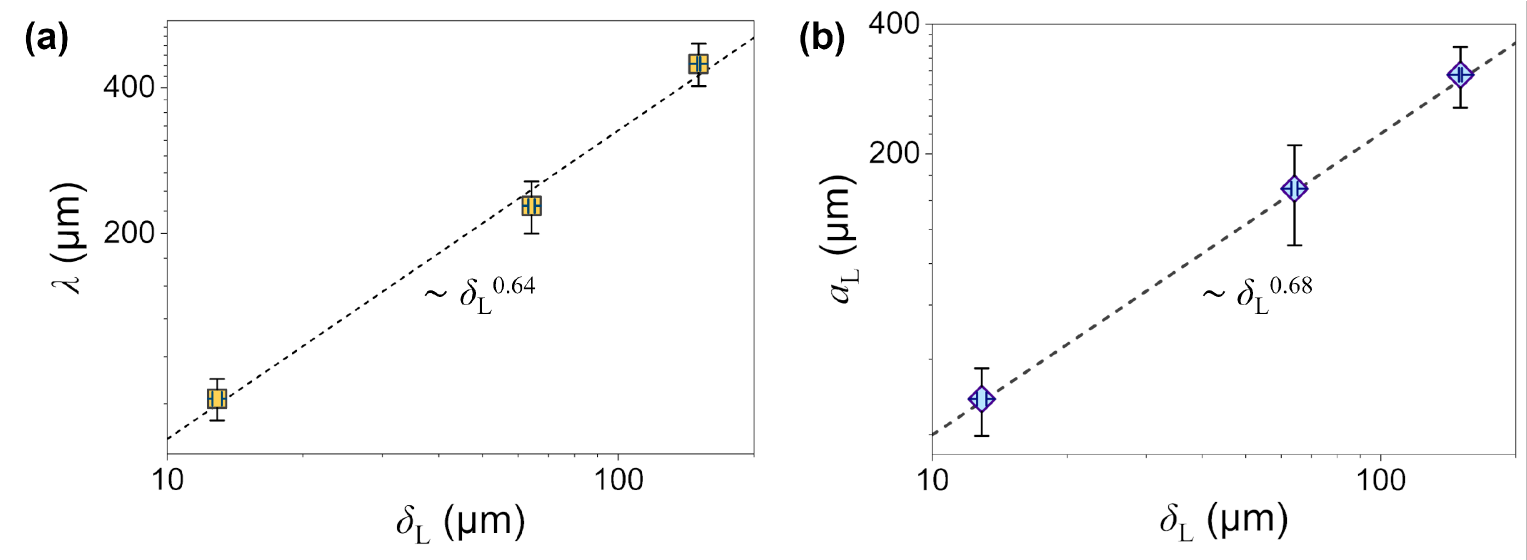}
\caption{Variation of experimentally measured (a) pattern wavelength $\lambda$ and (b) arm length $\aL$ with final film thickness $\dL$ for PDMS ($\Psix$), swollen in hexane followed by air drying. The dashed lines show the power-law best-fit curves, i.e., $\lambda\sim\dL^{0.64}$ and $\aL\sim\dL^{0.68}$.}
\label{fig2}
\end{figure}

In the present experiments, PDMS ($\Psix$) films with initial thicknesses of 30, 120, and \SI{1150}{\micro\metre}, all extracted for \SI{96}{\hour} and dried in air, yield final thicknesses $\dL\simeq13$, 64, and \SI{150}{\micro\metre}. In Fig.~\ref{fig2}, we show plots of wavelength $\lambda$ and arm length $\aL$ versus final thickness $\dL$. Independent power-law fits yield $\lambda\sim\dL^{0.64}$ and $\aL\sim\dL^{0.68}$, close to the exponent $5/8=0.625$, with the prefactors 17.3 and 8.8, respectively (Fig.~\ref{fig2}). Using $D\sim\SI{e-11}{\metre\squared\per\second}$, $\ve\sim\SI{1}{\micro\metre\per\second}$, and $\Es/\Ef\sim10^2$ in Eq.~(\ref{eq:lambda}) yields a prefactor of 7.5 for $\lambda\sim\dL^{0.64}$ in close agreement with our experimental data [Fig.~\ref{figS6}, Supplemental Material]. Notably, a factor of $(2\pi)^{1/2}$~\cite{Huang2005} separates the two. Independently performed AFM measurements and subsequent analysis reveal a skin stiffness of approximately \SI{500}{\kilo\pascal} for $\Psix$ films swollen in hexane and dried [Fig.~\ref{figS7} and Note S2, Supplemental Material]. Thus, our assumption of $\Es/\Ef\sim10^2$ is reasonable compared to the experimentally measured ratio, i.e., $\Es/\Ef=519/9.6\approx50$.

It is noteworthy to mention here that the observed 5/8 exponent is higher than both the 1/2 exponent predicted for an incompressible foundation~\cite{Chen2004,Huang2005} and the 1/4 exponent predicted by a compressible Winkler estimate with an undrained film layer~\cite{Huang2005}. However, it is lower than the linear scaling law observed by Tanaka \textit{et al.} for substrate-bounded swollen homogeneous gels without a skin layer~\cite{Tanaka1987}. Further, from Eq.~(\ref{eq:lambda}), we observe that the scaling law depends only on the stiffness ratio $\Es/\Ef$ rather than the absolute stiffness. Since this ratio is generated by the same densification of the skin in every formulation, the mesh size should ideally be insensitive to the modulus of the starting material. Experiments performed with additional PDMS formulations (PDMS 64:1, $P_{64}$, $E=\SI{4}{\kilo\pascal}$, and PDMS 80:1, $P_{80}$, $E=\SI{1}{\kilo\pascal}$), cast at a similar initial thickness, give wavelengths of \SI{89.3 \pm 10.3}{\micro\metre} and \SI{108.8 \pm 13.6}{\micro\metre}, and arm lengths of \SI{57.6 \pm 5.8}{\micro\metre} and \SI{59.4 \pm 9.0}{\micro\metre} [Fig.~\ref{figS8}, Supplemental Material], compared to $\lambda=\SI{91.0 \pm 8.9}{\micro\metre}$ and $\aL=\SI{53.8 \pm 9.7}{\micro\metre}$ for $\Psix$. A nearly tenfold change in modulus therefore shifts the arm length by only about 10\% and the wavelength by about 20\%. By contrast, any selection rule built on an absolute elastic length, such as $\gamma/E$ or $(\gamma\delta/E)^{1/2}$, would predict a three- to ten-fold change over the same range of moduli. The weak variation in pattern features, therefore, indicates that length selection is governed by kinematics rather than elasticity.

\begin{figure}[tb]
\includegraphics[width=\columnwidth]{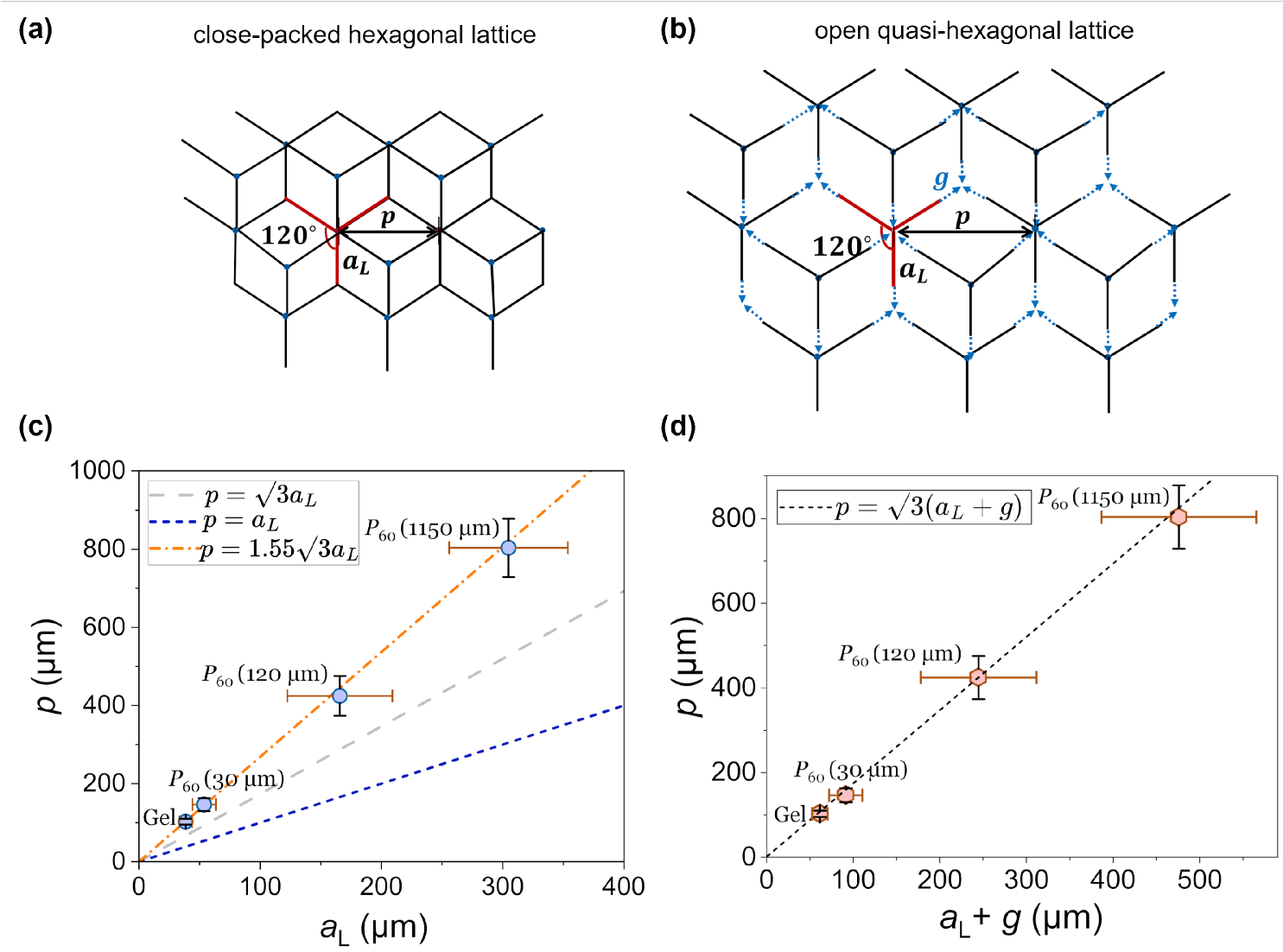}
\caption{(a) Ideal planform geometry. Vertices occupy a triangular lattice of pitch $p$. Each emits three arms at $120^\circ$ that terminate, a distance $p/\sqrt{3}$ away, on the centroids of adjacent lattice triangles, where they meet the arms of neighbouring vertices. The union is a close-packed hexagonal (honeycomb) lattice, and $p=\sqrt{3}\,\aL$. (b) Observed planform geometry: an open, quasi-hexagonal lattice in which each arm stops a gap $g$ short of the junction it would share with the arms of neighbouring vertices. (c) Measured pitch versus arm length for $\Psix$ films of initial thickness 30, 120, and \SI{1150}{\micro\metre} and for the silicone gel of around \SI{30}{\micro\metre} initial thickness, compared with the ideal relation $p=\sqrt{3}\,\aL$ (grey dashed line) and with $p=\aL$ (blue dashed line). The dash-dotted line is the best fit, $p=1.55\sqrt{3}\,\aL$. (d) Measured pitch versus $\aL+g$, compared with $p=\sqrt{3}(\aL+g)$ (dashed line).}
\label{fig3}
\end{figure}

\textbf{\textit{Why three arms, and why $\bm{120^\circ}$?}} Just above the instability threshold, the free surface and the pinned support break up-down symmetry, so the leading nonlinearity in the amplitude equation for the surface height is quadratic. A quadratic term resonantly couples wave-vector triads summing to zero, and among these the equilateral triad at $120^\circ$ is selected, making a hexagonal planform generic~\cite{Cross1993,Ciliberto1990}. The sign of the quadratic coefficient resolves the phase, setting the observed outward-protruding ridges~\cite{Zhao2025}. Far above onset, the deformation has sharpened into localized folds~\cite{Pocivavsek2008}, and the surface is better described as a network of fold lines, each with an energy per unit length. Mechanical equilibrium at a junction then obeys Plateau's rule, i.e., three lines of equal tension meeting at $120^\circ$, with any fourfold vertex unstable to dissociation into a pair of threefold ones~\cite{Weaire1999}. The weakly nonlinear and the fully localized limits thus converge on the same trivalent network. Accordingly, the $120^\circ$ angle is the single most robust feature of our experimental observations, invariant across both chemistries, all thicknesses, extraction times, and drying protocols [Fig.~\ref{fig1} and Figs.~\ref{figS3}, \ref{figS5}, \ref{figS7}--\ref{figS9}, Supplemental Material].

\begin{figure}[b]
\includegraphics[width=\columnwidth]{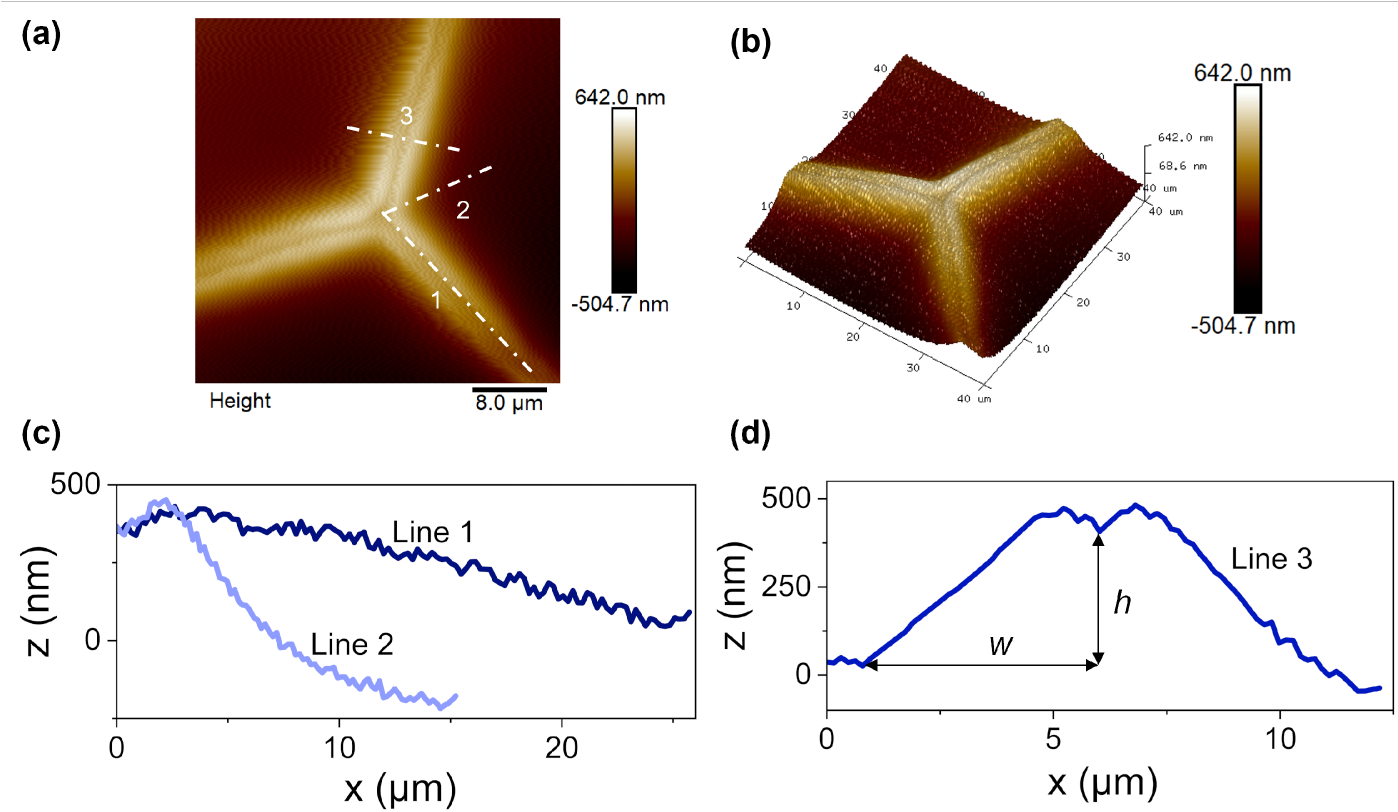}
\caption{(a) and (b) Atomic force microscopy scans of the pattern formation on $\Psix$ substrates swollen in hexane for 96 hr, followed by drying in air. (c) and (d) Profiles along the different dashed lines shown in (a). Representative ridge height $h$ and base half-width $w$ are shown for clarity. The average $h$ and $w$ are calculated using multiple such sectional scans.}
\label{fig4}
\end{figure}

Trivalency at $120^\circ$ is only one aspect of the network geometry, the other being the relation between the lattice pitch $p$ and the arm length $\aL$. In an ideal honeycomb, the arm length equals the bond length, so that $\aL=p/\sqrt{3}$ [Fig.~\ref{fig3}(a)]. Our experiments, however, reveal that the arms do not meet tip to tip but stop short of the junctions, leaving a gap [Fig.~\ref{fig3}(b) and Fig.~\ref{figS4}, Supplemental Material]. We therefore measure, in addition to the average pitch $p$, the average gap $g$ between the end point of an arm and the junction it would share with the arms of neighbouring vertices. Figure~\ref{fig3}(c) shows the variation of $p$ with $\aL$ for PDMS films of different thicknesses and for the silicone gel. The best fit gives $p\approx1.55\sqrt{3}\,\aL$, which exceeds the ideal honeycomb value by a factor of 1.55. When the gap is taken into account, however, the data collapses onto the relation $p=\sqrt{3}(\aL+g)$ [Fig.~\ref{fig3}(d)], as expected for an open honeycomb.

\textbf{\textit{Ridge amplitude.}} Although the skin is compressed by a strain of order 0.2, the folds accommodate only a minute fraction of this compression. AFM measurements reveal an average ridge height $h=\SI{0.42 \pm 0.20}{\micro\metre}$ and base half-width $w=\SI{5.63 \pm 0.92}{\micro\metre}$ (Fig.~\ref{fig4}). Approximating each ridge by a raised-cosine profile, which stores an excess arc length $\Delta\ell\simeq\pi^2h^2/(8w)$ per unit length~\cite{Cerda2003,Pocivavsek2008}, and noting that the open network carries a fold length per unit area $\rho_\mathrm{f}=(2/\sqrt{3})\,\aL/(\aL+g)^2$ distributed equally over three orientations, the network relieves an equibiaxial in-plane strain (Note S4, Supplemental Material)
\begin{equation}
\varepsilon_\mathrm{f} = \tfrac{1}{2}\rho_\mathrm{f}\Delta\ell = \frac{\pi^2}{8\sqrt{3}}\,\frac{h}{w}\,\frac{h}{\aL}\left(\frac{\aL}{\aL+g}\right)^{2}.
\label{eq:epsf}
\end{equation}

Equation~(\ref{eq:epsf}) shows that the absorbed strain is the product of the ridge slope $h/w$, the ratio $h/\aL$ of ridge height to arm length, and a geometric filling factor $[\aL/(\aL+g)]^2=3\aL^2/p^2\leq1$ that accounts for the unfolded gaps. Using $\aL\simeq\SI{54}{\micro\metre}$ and $p\simeq1.55\sqrt{3}\,\aL$ [Fig.~\ref{fig3}(c)], Eq.~(\ref{eq:epsf}) gives $\varepsilon_\mathrm{f}\simeq1.73\times10^{-4}$, less than 0.1\% of the available compression, and absorbing the full compression would require ridges around forty times taller than measured (Note S4, Supplemental Material). Unlike the folds of a laterally confined elastic sheet, which grow until they consume the imposed displacement~\cite{Cerda2003,Pocivavsek2008}, the folds here are arrested long before the excess area is consumed, while the residual mismatch remains stored in the skin and relaxes slowly thereafter. The amplitude is consequently set not by the available strain but by the competition between surface stress and elasticity at the elastocapillary length~\cite{Style2017,Pandey2020}, $h\sim\ell_\mathrm{EC}=\gamma/E$. As the network densifies, its modulus rises from the swollen value by up to $(1/\varphi_\mathrm{p})^{n}$, with $1\leq n\leq7/3$~\cite{deGennes1979}, so that $\ell_\mathrm{EC}$ spans 0.26 to \SI{2.2}{\micro\metre} for the PDMS ($\Psix$) against $h\simeq\SI{0.42}{\micro\metre}$, and 0.35 to \SI{3.8}{\micro\metre} for the gel against $h\simeq\SI{2.5}{\micro\metre}$ [Fig.~\ref{figS10} and Note S4, Supplemental Material], with $\gamma\simeq\SI{21}{\milli\newton\per\metre}$~\cite{Style2017}.

In conclusion, the triradial motifs emerging in solvent-extracted viscoelastic thin films and gels are not isolated defects, but interconnected vertices of an ordered honeycomb fold network whose length scales are governed by distinct physical mechanisms. First, the in-plane network geometry imposes $p/(\aL+g)=\sqrt{3}$, obeyed by two chemically distinct networks. Second, network spacing is governed by the kinetics of solvent removal: a diffusively growing skin bending over a drained, compressible underlayer yields $\lambda\sim(\Es/\Ef)^{1/4}(D/\ve)^{3/8}\dL^{5/8}$, a scaling that also captures the arm length despite a tenfold difference in modulus. Finally, the folds accommodate a strain $\varepsilon_\mathrm{f}\approx1.73\times10^{-4}$, nearly three orders of magnitude smaller than the compression driving their formation, and their height is set by elastocapillary arrest at the length $\gamma/E$ of the locally densified network. Network morphology therefore encodes the kinetics of solvent extraction rather than the bulk stiffness of the material, making film thickness and drying rate independent, largely modulus-insensitive control parameters for programming ordered surface topographies in soft materials.

\begin{acknowledgments}
\textit{Acknowledgments}---F.Z. is supported by the Waterloo Institute for Nanotechnology-Soochow Doctoral Program Scholarship. S.K.M. acknowledges support from Discovery Grant No.~RGPIN-2024-03729 of the Natural Sciences and Engineering Research Council of Canada. B.Z. acknowledges support from NSERC Grant No.~RGPIN-2019-04650. Z.L. acknowledges support from NSERC Grant No.~RGPIN-50503-11497 and the Canadian Foundation for Innovation, CFI-JELF Grant No.~52479-10255. The authors used Claude (Anthropic, v4.8 and v5.5) to assist in checking the scaling estimates and in editing the manuscript; the authors reviewed all content and take full responsibility for it.

\textit{Author contributions}---S.K.M. conceived the research. F.Z. performed all the experiments with assistance from S.M. F.Z., S.M., B.Z., and S.K.M. analyzed and discussed the results. S.T. and Z.L. performed the AFM and force spectroscopy measurements. S.M. wrote the manuscript with input from all the authors. B.Z. and S.K.M. supervised the research.
\end{acknowledgments}

\textit{Data availability}---The data that support the findings of this article are available from the authors upon reasonable request.

\clearpage
\onecolumngrid

\setcounter{equation}{0}
\setcounter{figure}{0}
\setcounter{table}{0}
\renewcommand{\theequation}{S\arabic{equation}}
\renewcommand{\thefigure}{S\arabic{figure}}
\renewcommand{\thetable}{S\arabic{table}}

\begin{center}
\textbf{\large Supplemental Material for ``Drying kinematics selects open honeycomb fold networks on solvent-extracted viscoelastic thin films and gels''}
\end{center}
\vspace{1em}

\noindent\textbf{Contents}
\begin{quote}
Note S1. Materials and methods\\
Note S2. AFM imaging and nanoindentation\\
Note S3. Wavelength selection by a growing skin on a drained foundation [Eqs.~(\ref{eq:skin}) and (\ref{eq:lambda})]\\
Note S4. Strain absorbed by the fold network and ridge amplitude [Eq.~(\ref{eq:epsf})]\\
Tables \ref{tabS1}--\ref{tabS2}; Figures \ref{figS1}--\ref{figS10}
\end{quote}

\section*{Note S1. Materials and methods}

\textit{Materials.}---We prepared PDMS films from Sylgard 184 (Dow), consisting of a base and a curing agent. We prepared the silicone gel from a vinyl-terminated base (Gelest DMS-V31), a platinum catalyst, and a hydride curing agent~\cite{Zhao2025}. We used hexane, ethanol, and isopropanol as received.

\textit{Film fabrication.}---PDMS films $\Psix$, $P_{64}$, and $P_{80}$ were prepared by mixing base and curing agent at weight ratios of 60:1, 64:1, and 80:1, mixing and degassing in a planetary mixer (Thinky ARE-310) for \SI{5}{\minute}, and spin coating/drop casting onto glass slides ($38.1\times25.4\times\SI{1}{\milli\metre\cubed}$) that had been rinsed with isopropanol and ethanol and dried under nitrogen. Films were cured at \SI{80}{\celsius} for \SI{14}{\hour}. We prepared the gel by mixing the base and catalyst (10:0.005 by weight) and, separately, the base and curing agent (10:1), then combining the two at a 9:1 ratio, degassing, spin-coating, and curing at \SI{80}{\celsius} for \SI{14}{\hour}~\cite{Zhao2025}. As control substrates (Fig.~\ref{figS1}), $\Psix$ was spin-coated onto 0.5-mm-thick cured $P_{10}$ (PDMS, 10:1) substrates and onto bare silicon wafers and cured under the same conditions.

\textit{Extraction and drying.}---Films were placed coated-side up in a beaker, and \SI{200}{\milli\litre} of hexane was gently introduced along the wall to avoid disturbing the surface. The beaker was covered with aluminum foil to limit evaporation. Extraction times were $\tE=\SI{8}{\minute}$, \SI{1}{\hour}, \SI{10}{\hour}, and \SI{96}{\hour}, with the hexane refreshed every \SI{48}{\hour} for the longest. Films were then dried in ambient air unless stated otherwise. Drying for \SI{10}{\minute} or \SI{12}{\hour} in air, \SI{12}{\hour} in a fume hood, or \SI{12}{\hour} in an oven (\SI{60}{\celsius} for the gel, \SI{80}{\celsius} for the PDMS) caused minor variations in the ridge height and width but not the mesh [Fig.~\ref{figS9} and Ref.~\cite{Zhao2025}].

\textit{Mass loss and thickness.}---The mass loss $\varphi$ was obtained by weighing each film before extraction and after extraction and drying. Since the extracted free chains and the network are chemically identical, mass and volume fractions coincide within the density difference between them, which we neglect. Film thicknesses were measured from cross-sections by field-emission scanning electron microscopy (Zeiss UltraPlus)~\cite{Zhao2025}.

\textit{Interference microscopy.}---Dual-wavelength reflection interference contrast microscopy (DW-RICM) was performed on an inverted Zeiss LSM 800 with a $10\times$ objective (EC Plan-Neofluar, N.A.~$=0.3$) at \SI{488}{\nano\metre} (represented as green channel images) and \SI{561}{\nano\metre} (represented as red channel images).

\textit{Pattern metrology.}---For each film, $\lambda$, $\aL$, $p$, and $g$ were measured on $N=150$--250 motifs from DW-RICM images. The pattern wavelength $\lambda$ was measured as the average of nearest-neighbour vertices; the fold arm length $\aL$ from the vertex to the arm tip; the pitch $p$ between like-oriented vertices; and the gap $g$ from an arm tip to the point where the three converging arms would meet.

\section*{Note S2. AFM imaging and nanoindentation}

We performed AFM measurements on solvent-extracted, air-dried $\Psix$ films on glass. Images were acquired in ac (non-contact) mode on a Bruker JPK NanoWizard IV with QP-BioAC-CB2 cantilevers driven at their nominal resonance frequency. Force spectroscopy was then performed in contact mode on the same area with the same cantilever, after recalibrating the spring constant against the manufacturer's value. Points were selected on and off the ridges, and 100 force--distance curves were recorded at each point, giving $N=200$ curves per sample. Young's moduli were extracted with the JPK data-processing software using the Derjaguin--Muller--Toporov (DMT) model~\cite{DMT1975} for a paraboloidal tip of radius $R=\SI{10}{\nano\metre}$,
\begin{equation}
F + F_\mathrm{adh} = \frac{4}{3}\,\frac{E}{1-\nu^{2}}\,R^{1/2}\,d^{3/2},
\label{eq:DMT}
\end{equation}
where $F$ is the applied load, $F_\mathrm{adh}$ is the pull-off (adhesion) force, $d$ is the indentation depth, and $\nu$ is Poisson's ratio.

The surface modulus is \SI{519 \pm 71}{\kilo\pascal} on the ridges and \SI{315 \pm 27}{\kilo\pascal} between them (mean $\pm$ s.d.; Fig.~\ref{figS7}). Both values exceed the as-cast modulus of $\Psix$ (\SI{9.6}{\kilo\pascal}) by a factor of 30 to 50, confirming that the whole surface carries a densified skin, and the ridges are about 65\% stiffer than the flat regions between them.

\section*{Note S3. Wavelength selection by a growing skin on a drained foundation [Eqs.~(2) and (3)]}

\textit{Why the skin is not a solvent boundary layer.}---Hexane diffuses through swollen PDMS with $D_\mathrm{f}\sim\SI{e-9}{\metre\squared\per\second}$~\cite{Hu2011,Phadnis2018}. For a film receding at $\ve\sim\SI{1}{\micro\metre\per\second}$, the solvent P\'eclet number $\ve\dL/D_\mathrm{f}$ is 0.013, 0.064, and 0.15 for dried film thicknesses of $\dL=12.9$, 64, and \SI{150}{\micro\metre}. Thus, solvent gradients across the film are weak. Instead, a skin forms because the polymer-rich surface layer relaxes and redistributes much more slowly, with an effective diffusivity $D_\mathrm{s}\sim\SI{e-11}{\metre\squared\per\second}$ of the order of the self-diffusivity of the mobile chains~\cite{Appel1993}. The skin P\'eclet number $\ve\dL/D_\mathrm{s}$ is then 1.3, 6.4, and 15, indicating formation of the skin. Once it limits evaporation, it thickens diffusively throughout drying~\cite{Okuzono2006,Talini2023}, giving Eq.~(\ref{eq:skin}) of the main text, $\ls\sim(D_\mathrm{s}\dL/\ve)^{1/2}$. A steady evaporation boundary layer of fixed thickness $\ls\sim D_\mathrm{s}/\ve$ would instead give $\lambda\sim\dL^{1/4}$ (Table~\ref{tabS2}), which significantly deviates from the measured exponent.

\textit{Plate on a foundation.}---Let the skin, of bending stiffness $B=\Es\ls^3/12$ (plane-strain modulus), carry an in-plane compressive force per unit length $N$ and rest on a foundation of stiffness $K(q)$. For a deflection $\zeta=A\cos qx$, the energy per unit area is
\begin{equation}
U = \frac{A^{2}}{4}\left[Bq^{4} + K(q) - Nq^{2}\right],
\label{eq:U}
\end{equation}
so the flat state is unstable once $N$ exceeds $\min_q[Bq^2+K(q)/q^2]$~\cite{Cerda2003,Hetenyi1946}.

\textit{Drained foundation.}---Since the underlying film can expel liquid during the drying time, its normal response is governed by its drained (oedometric) modulus $M=\Ef(1-\nu_\mathrm{d})/[(1+\nu_\mathrm{d})(1-2\nu_\mathrm{d})]\sim\Ef$ rather than by incompressibility. For $q\dL\lesssim1$, the response is local, $K=M/\dL\sim\Ef/\dL$~\cite{Huang2005,Brau2013}. Minimizing gives
\begin{equation}
q^{*} = \left(\frac{K}{B}\right)^{1/4},\qquad
\lambda = 2\pi\left(\frac{B}{K}\right)^{1/4} = 2\pi\left(\frac{\Es\ls^{3}\dL}{12\Ef}\right)^{1/4},\qquad
N_\mathrm{c} = 2\sqrt{BK},
\label{eq:drained}
\end{equation}
and substituting Eq.~(\ref{eq:skin}) in the main text provides Eq.~(\ref{eq:lambda}) (main text), $\lambda\sim(\Es/\Ef)^{1/4}(D_\mathrm{s}/\ve)^{3/8}\dL^{5/8}$. The corresponding elastic threshold strain is $\varepsilon_\mathrm{th}=N_\mathrm{c}/(\Es\ls)=(\ls/\dL)^{1/2}[\Ef/(3\Es)]^{1/2}$. Using $\Es=\SIrange{300}{500}{\kilo\pascal}$ (from AFM) and obtaining $\ls$ from Eq.~(\ref{eq:skin}), $\varepsilon_\mathrm{th}=0.04$--0.10, i.e., below the available strain $\varepsilon\approx0.2$--0.26.

\textit{Incompressible foundation.}---If instead the underlying film could not change volume, normal deflection would require lateral flow, and lubrication yields $K(q)\simeq cG_\mathrm{f}/(\dL^3q^2)$, with $c=3$ for a shear-free top surface and $c=12$ for a no-slip one. Minimizing $Bq^2+K(q)/q^2$ then yields $\lambda\sim(B\dL^3/G_\mathrm{f})^{1/6}\sim(\ls\dL)^{1/2}$.

Table~\ref{tabS2} exhibits the exponents that follow from each combination of skin and foundation.

\section*{Note S4. Strain absorbed by the fold network and ridge amplitude [Eq.~(4)]}

\textit{Ridge profile.}---We approximate the cross section of a ridge of height $h$ and base half-width $w$ by a raised-cosine profile, $z(x)=(h/2)[1+\cos(\pi x/w)]$ for $|x|\leq w$. Representative AFM measurements reveal an average ridge height $h=\SI{0.42 \pm 0.20}{\micro\metre}$ and an average half-width $w=\SI{5.63 \pm 0.92}{\micro\metre}$ (Fig.~\ref{fig4}, main text). In the small-slope limit, the excess arc length that such a profile stores per unit length along its axis is $\Delta\ell\simeq\pi^2h^2/(8w)=\SI{0.039}{\micro\metre}$~\cite{Cerda2003,Pocivavsek2008}, an approximation that is accurate to 1\% for the measured average ridge slope $h/w\approx0.074$.

\textit{Fold density of the open network.}---Because the arms terminate a distance $g$ short of the junctions [Fig.~\ref{fig3}(b), main text], the fold network is open rather than a closed honeycomb. Each cell of the triangular lattice, of area $(\sqrt{3}/2)\,p^2$, contains a single triradial carrying a fold length $3\aL$, so that the fold length per unit area is $\rho_\mathrm{f}=2\sqrt{3}\,\aL/p^2=(2/\sqrt{3})\,\aL/(\aL+g)^2$, which reduces to the closed-honeycomb value $2/(\sqrt{3}\,\aL)$ when $g=0$.

\textit{Strain relieved by the network.}---Since a fold takes up excess length only in the direction normal to its own axis~\cite{Cerda2003}, a fold with in-plane unit normal $\nhat$ relieves a strain proportional to $\nhat\nhat$. Because all triradials share the same orientation, the arms are distributed equally among three orientations separated by $60^\circ$, for which $\langle\nhat\nhat\rangle=\mathbf{I}/2$, and the network as a whole therefore relieves an equibiaxial in-plane strain of magnitude $\varepsilon_\mathrm{f}=\tfrac{1}{2}\rho_\mathrm{f}\Delta\ell$, which gives Eq.~(\ref{eq:epsf}) of the main text. The particular choice of profile affects only the numerical prefactor; a triangular ridge, for example, which minimizes $\Delta\ell$ for given $h$ and $w$, replaces $\pi^2/8$ by unity.

\textit{Comparison with the available compression.}---Substituting $h\simeq\SI{0.42}{\micro\metre}$, $w\simeq\SI{5.6}{\micro\metre}$, and $\aL\simeq\SI{54}{\micro\metre}$, together with the measured relation $p\simeq1.55\sqrt{3}\,\aL$ [Fig.~\ref{fig3}(c), main text], for which the filling factor is $1/1.55^2\approx0.42$, gives $\varepsilon_\mathrm{f}\simeq1.73\times10^{-4}$. This amounts to less than 0.1\% of the available compression, whether it is compared with the upper bound $\varepsilon\simeq0.26$ obtained from Eq.~(\ref{eq:strain}) of the main text or with the onset value $\varepsilon_\mathrm{c}\simeq0.23$ [Fig.~\ref{fig1}(d), main text]. Conversely, if the folds were to absorb the full compression $\varepsilon$ at the observed values of $w$, $\aL$, and $p$, Eq.~(\ref{eq:epsf}) would require a ridge height $h_\varepsilon=[8w\varepsilon p^2/(\sqrt{3}\pi^2\aL)]^{1/2}\simeq\SI{16.3}{\micro\metre}$, around forty times larger than the measured height.

\textit{Elastocapillary range of the amplitude.}---The bulk modulus sweeps over orders of magnitude as the network densifies, from the swollen value to $(1/\varphi_\mathrm{p})^{n}$ times it, with $n$ between 1 for affine dilution and 7/3 for scaling~\cite{deGennes1979}. With $\gamma\simeq\SI{21}{\milli\newton\per\metre}$~\cite{Style2017}, the elastocapillary length $\ell_\mathrm{EC}=\gamma/E$ thus varies between 0.26 and \SI{2.2}{\micro\metre} for the elastomer, against $h\simeq\SI{0.42}{\micro\metre}$, and between 0.35 and \SI{3.8}{\micro\metre} for the gel, against $h\simeq\SI{2.5}{\micro\metre}$ (Fig.~\ref{figS10}).

\clearpage
\section*{Supplemental tables}

\begin{table}[h]
\caption{Mass loss $\varphi$ after extraction and drying, the corresponding estimate of the released strain from Eq.~(\ref{eq:strain}) with $\varphi$ as a proxy for $\Delta\varphi$, and whether a well-ordered triradial array is reported (Fig.~\ref{figS5} and Ref.~\cite{Zhao2025}).}
\label{tabS1}
\begin{ruledtabular}
\begin{tabular}{ccccccc}
$\tE$ & $\varphi$ ($\Psix$) & $\varepsilon$ ($\Psix$) & ordered ($\Psix$) & $\varphi$ (gel) & $\varepsilon$ (gel) & ordered (gel) \\
\hline
\SI{8}{\minute}  & 0.454 & 0.182 & no  & 0.510 & 0.211 & yes \\
\SI{1}{\hour}    & 0.537 & 0.226 & no  & 0.621 & 0.277 & yes \\
\SI{10}{\hour}   & 0.563 & 0.241 & yes & 0.634 & 0.285 & yes \\
\SI{96}{\hour}   & 0.600 & 0.263 & yes & 0.637 & 0.287 & yes \\
\end{tabular}
\end{ruledtabular}
\end{table}

\begin{table}[h]
\caption{Thickness exponent of the fold wavelength, $\lambda\sim\dL^{\beta}$, for each combination of skin and foundation (Note S3). The diffusively growing skin has $\ls\sim\dL^{1/2}$ [Eq.~(\ref{eq:skin})].}
\label{tabS2}
\begin{ruledtabular}
\begin{tabular}{llcc}
Skin & Foundation & Wavelength & $\beta$ \\
\hline
fixed thickness & drained (Winkler), $K\sim\Ef/\dL$ & $(\ls^3\dL)^{1/4}$ & 1/4 \\
fixed thickness & incompressible, $K\sim G_\mathrm{f}/(\dL^3q^2)$ & $(\ls\dL)^{1/2}$ & 1/2 \\
growing, $\ls\sim\dL^{1/2}$ & drained (Winkler) & $(\ls^3\dL)^{1/4}$ & 5/8 \\
growing, $\ls\sim\dL^{1/2}$ & incompressible & $(\ls\dL)^{1/2}$ & 3/4 \\
none (homogeneous swollen gel) & --- & $\sim\dL$~\cite{Tanaka1987} & 1 \\
measured, $\Psix$ & & $\lambda$; $\aL$ & 0.64; 0.68 \\
\end{tabular}
\end{ruledtabular}
\end{table}

\clearpage
\section*{Supplemental figures}

\begin{figure}[h]
\includegraphics[width=0.85\linewidth]{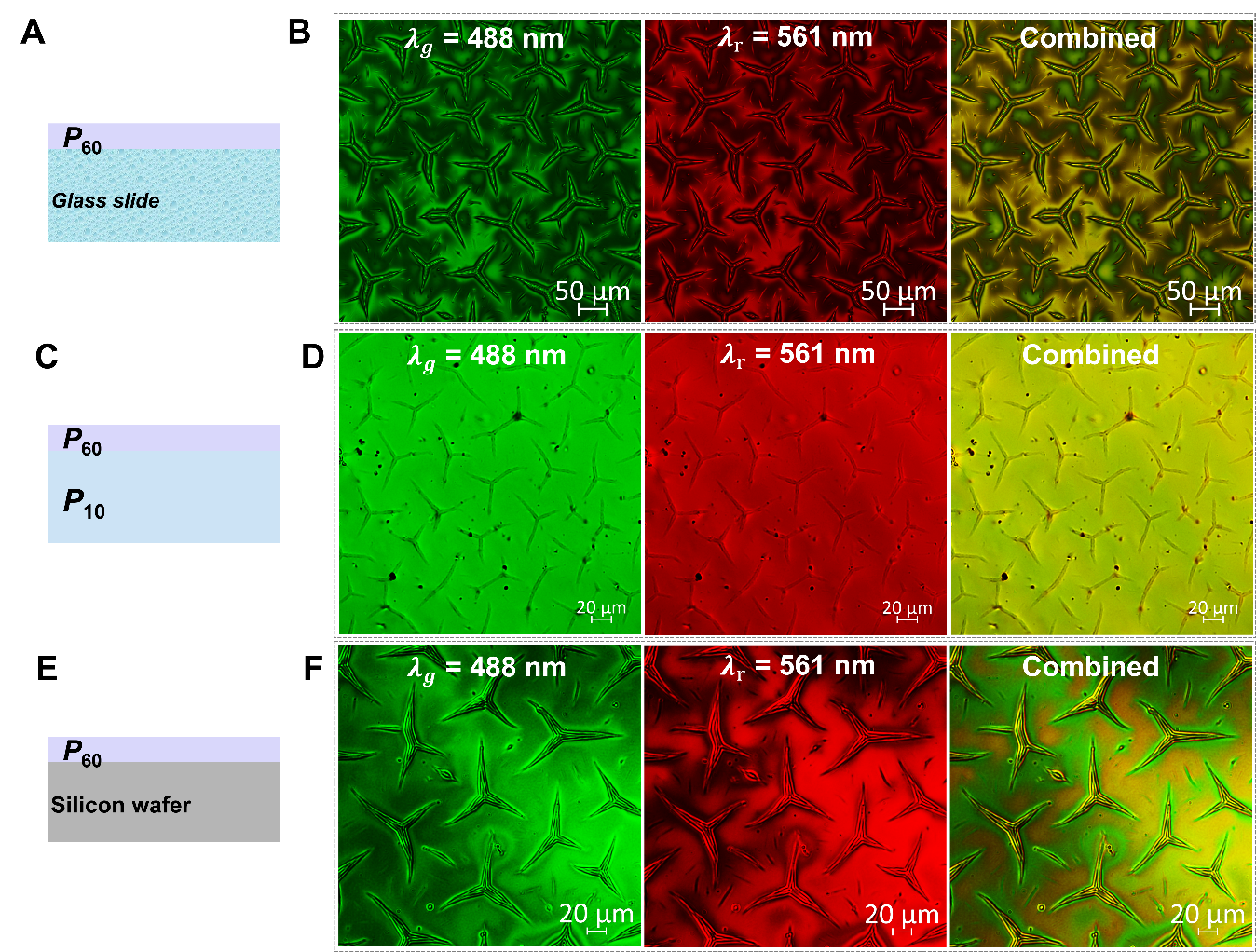}
\caption{(A) (B) and (C) Schematic of the $\Psix$ (PDMS base and curing agent ratio: 60:1) layer coated on a glass slide, a thicker $P_{10}$ (PDMS base and curing agent ratio: 10:1) substrate and a silicon wafer, respectively. (D) (E) and (F) Dual-wavelength reflection interference contrast microscopy (DW-RICM) images of the $\Psix$ thin film coated on a glass slide, a thicker $P_{10}$ (PDMS base and curing agent ratio: 10:1) substrate and a silicon wafer, respectively, with a hexane extraction time ($\tE$) of 96 hr and followed by air drying for 1 day. The green and red channels correspond to imaging using laser wavelengths of \SI{488}{\nano\metre} and \SI{561}{\nano\metre}, respectively. The yellow images correspond to the overlay of the green and red channels.}
\label{figS1}
\end{figure}

\begin{figure}[h]
\includegraphics[width=0.85\linewidth]{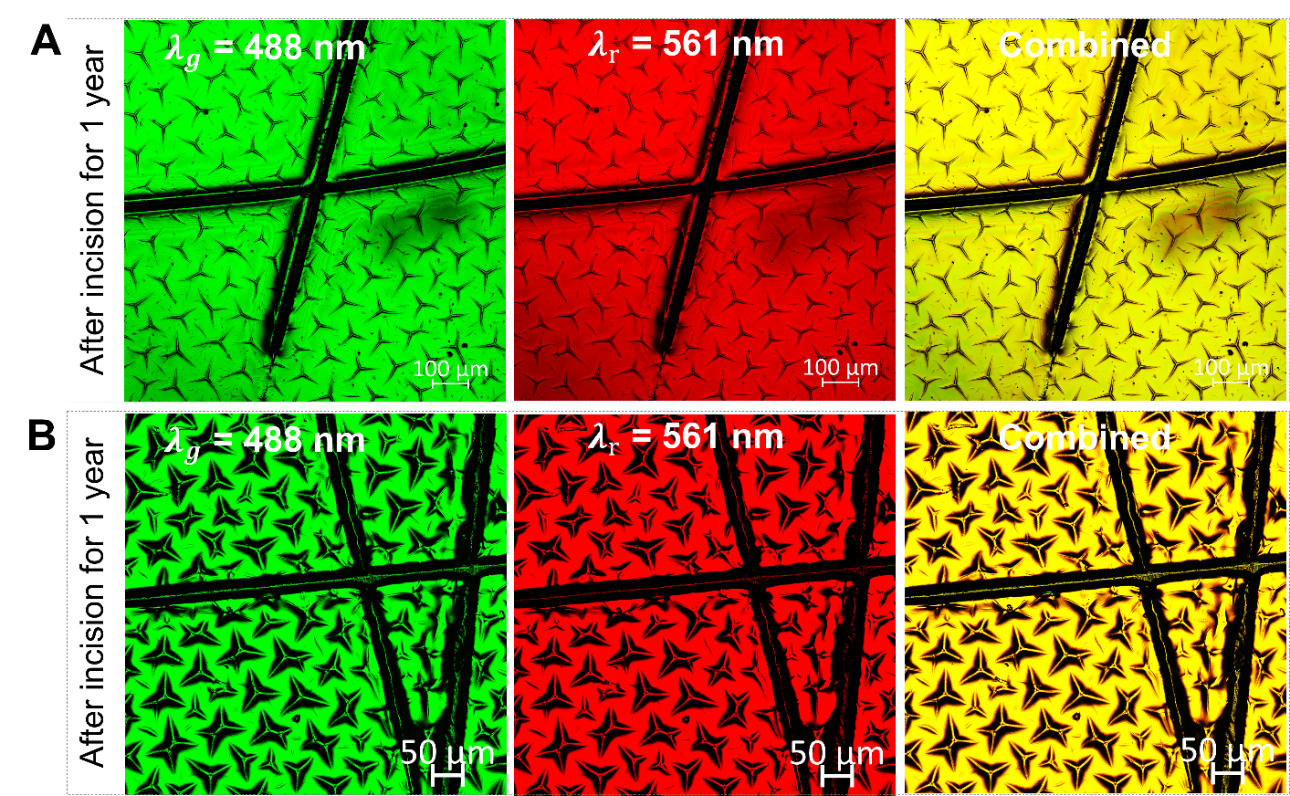}
\caption{Dual-wavelength reflection interference contrast microscopy (DW-RICM) images of (A) the soft PDMS $\Psix$ substrate and (B) the silicone gel substrate after 96 hours of hexane extraction and drying in air, and subsequent storage under laboratory conditions for one year. The images show the morphology of the surface pattern in a region where a small incision is made using a blade to examine pattern change across the incision. The green (first column) and red (second column) channels show images acquired at \SI{488}{\nano\metre} and \SI{561}{\nano\metre}, respectively. The combined channel image (third column) is the overlay of the green and red channel images.}
\label{figS2}
\end{figure}

\begin{figure}[h]
\includegraphics[width=0.85\linewidth]{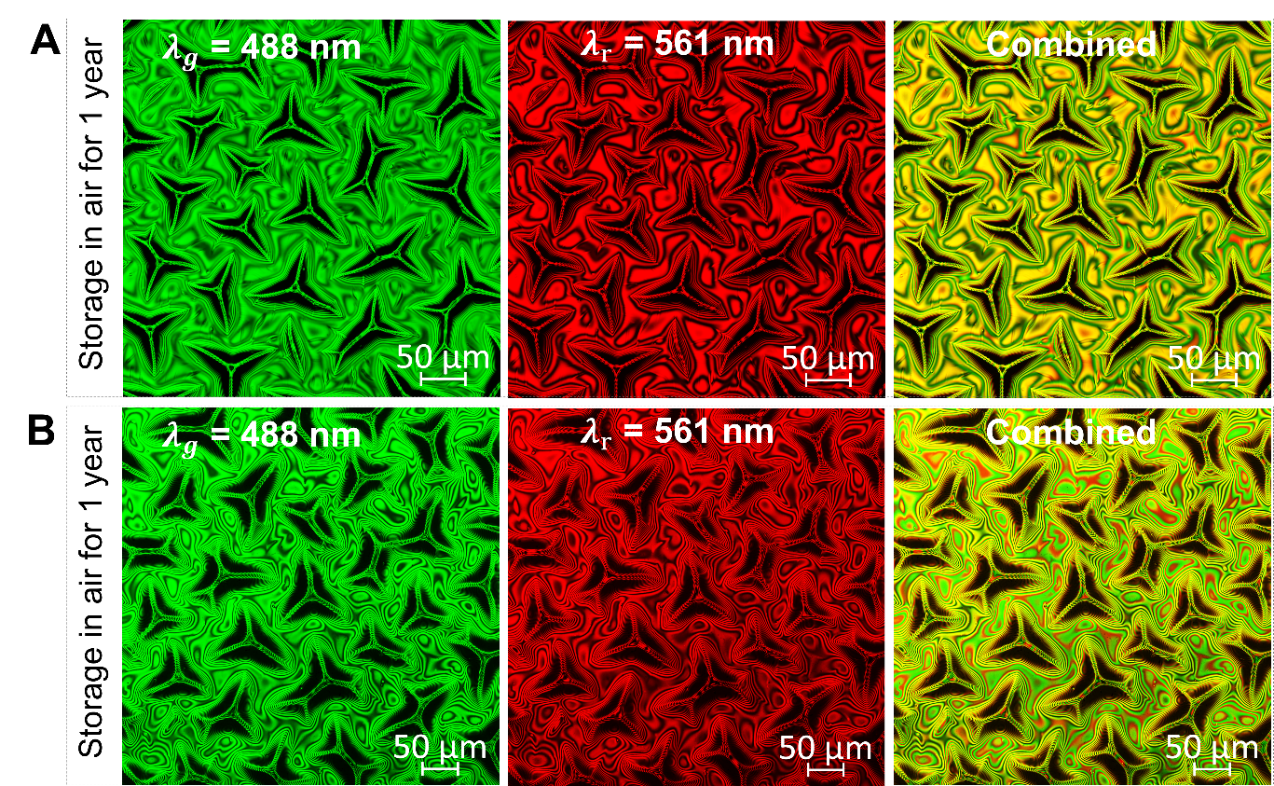}
\caption{Dual-wavelength reflection interference contrast microscopy (DW-RICM) images of the surface morphology of (A) $P_{64}$ (PDMS base and curing agent ratio: 64:1) substrate and (B) $P_{80}$ (PDMS base and curing agent ratio: 80:1) substrate after 96 hr of hexane extraction and air drying, and subsequent storage under laboratory conditions for one year. The green (first column) and red (second column) channels show images acquired at \SI{488}{\nano\metre} and \SI{561}{\nano\metre}, respectively. The combined channel image (third column) is the overlay of the green and red channel images.}
\label{figS3}
\end{figure}

\begin{figure}[h]
\includegraphics[width=0.42\linewidth]{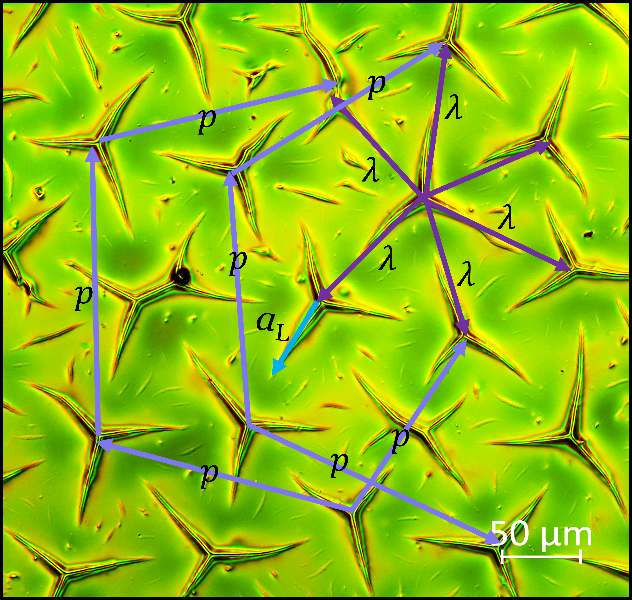}
\caption{DW-RICM images of surface patterns on $\Psix$ highlighting the measurement of the fold wavelength $\lambda$, the arm length $\aL$, the honeycomb pitch $p$, and the gap $g$.}
\label{figS4}
\end{figure}

\begin{figure}[h]
\includegraphics[width=0.85\linewidth]{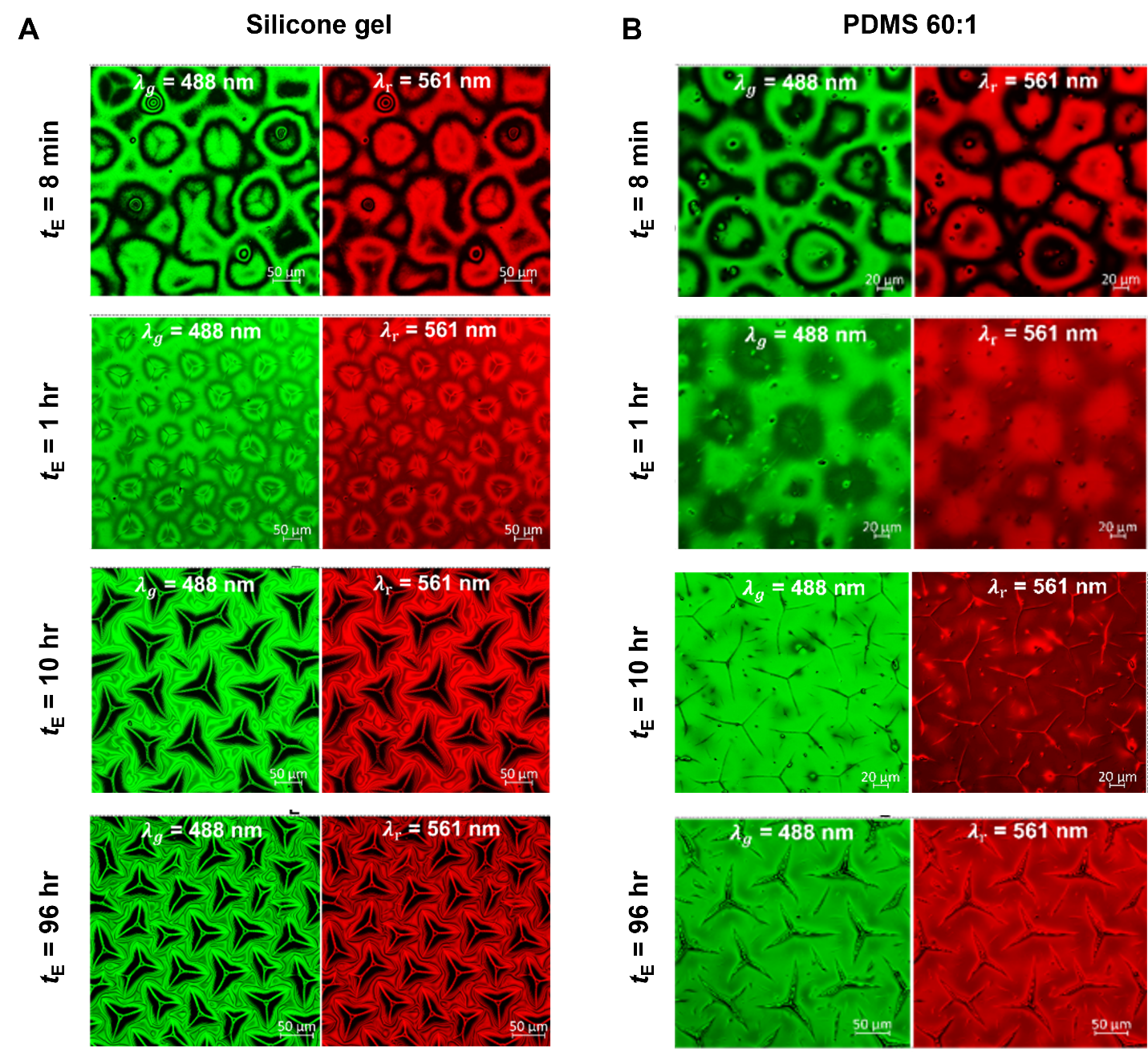}
\caption{Dual-wavelength reflection interference contrast microscopy (DW-RICM) images of surface morphology and pattern evolution for (A) silicone gel substrate and (B) PDMS (60:1) substrate for different hexane extraction times $\tE$, followed by air drying. The green (first column) and red (second column) channels show images acquired at \SI{488}{\nano\metre} and \SI{561}{\nano\metre}, respectively.}
\label{figS5}
\end{figure}

\begin{figure}[h]
\includegraphics[width=0.75\linewidth]{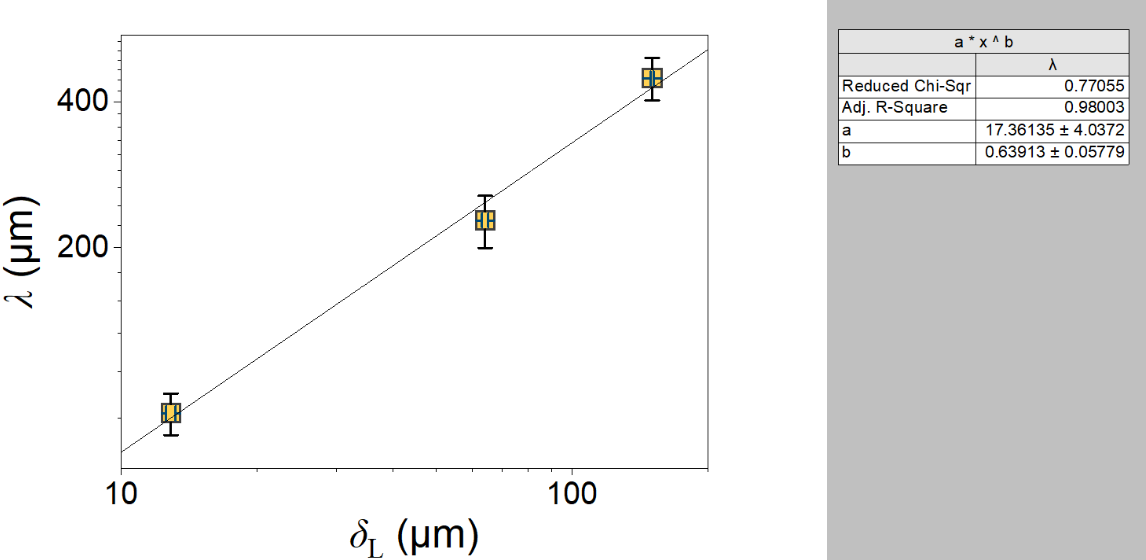}
\caption{Plot of variation of pattern wavelength $\lambda$ with final film thickness $\dL$ for PDMS ($\Psix$), swollen in hexane followed by air drying. Symbols represent experimental data, and the solid line represents the fitted curve. The power-law fitting expression is shown for clarity.}
\label{figS6}
\end{figure}

\begin{figure}[h]
\includegraphics[width=0.8\linewidth]{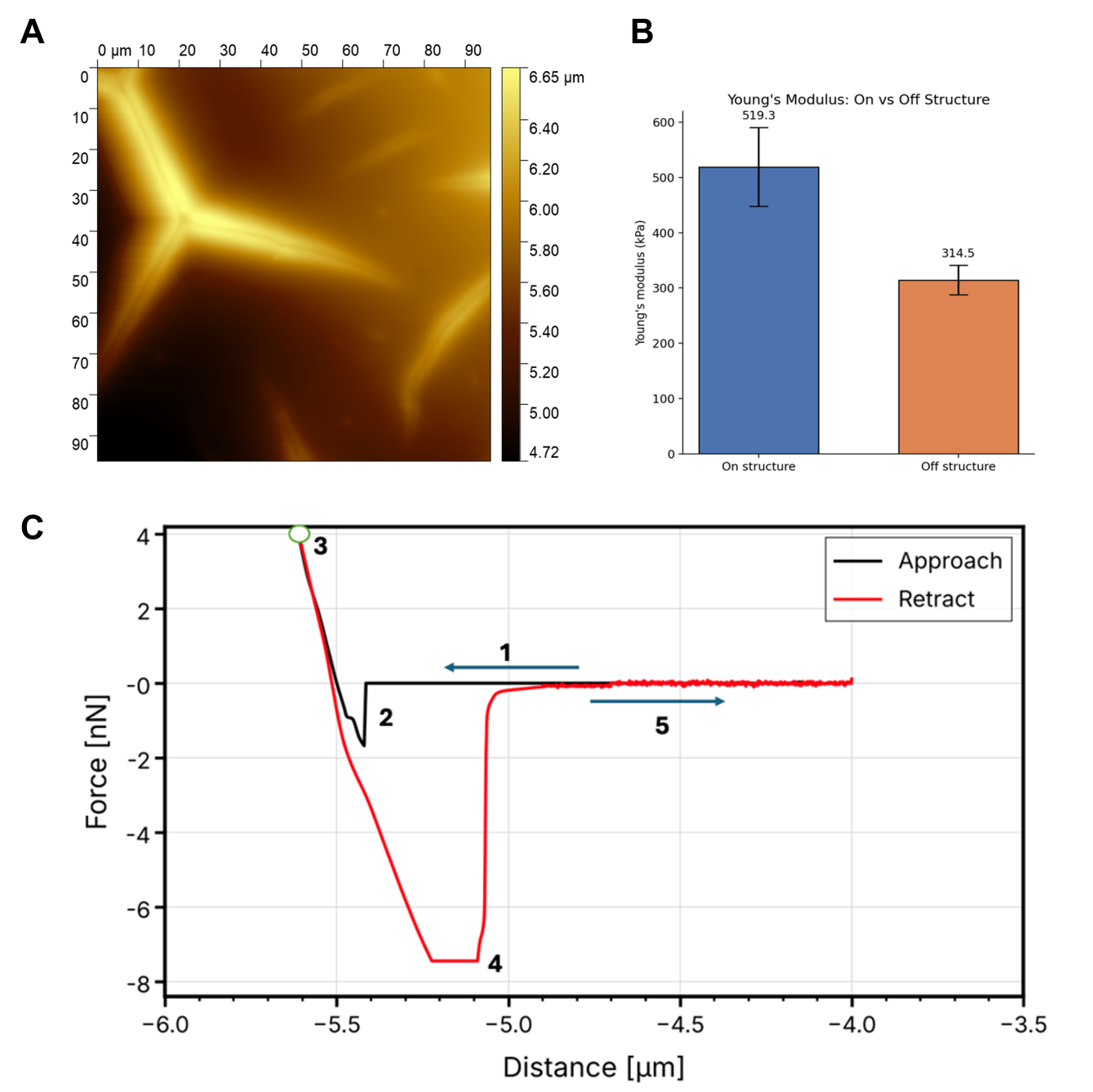}
\caption{(A) Representative AFM image ($100\times\SI{100}{\micro\metre}$) of patterns formed on PDMS film ($\Psix$) following hexane extraction and air drying. AFM imaging was done in air in non-contact mode. (B) Averaged Young's modulus measured ``on'' and ``off'' the pattern structures, with standard deviation. (C) Representative force-distance curve. 1) Tip approaching the sample surface from the retracted piezo position. 2) Tip jumps into contact with the sample during approach. 3) Repulsive force increases as the tip is pressed into the surface; the cantilever bends, and the force increases until it reaches the setpoint (circled in blue). 4) The piezo begins retracting the tip while the tip is still adhered to the surface. 5) The tip is pulled away from the surface and continues to retract until it reaches the initial piezo position.}
\label{figS7}
\end{figure}

\begin{figure}[h]
\includegraphics[width=0.85\linewidth]{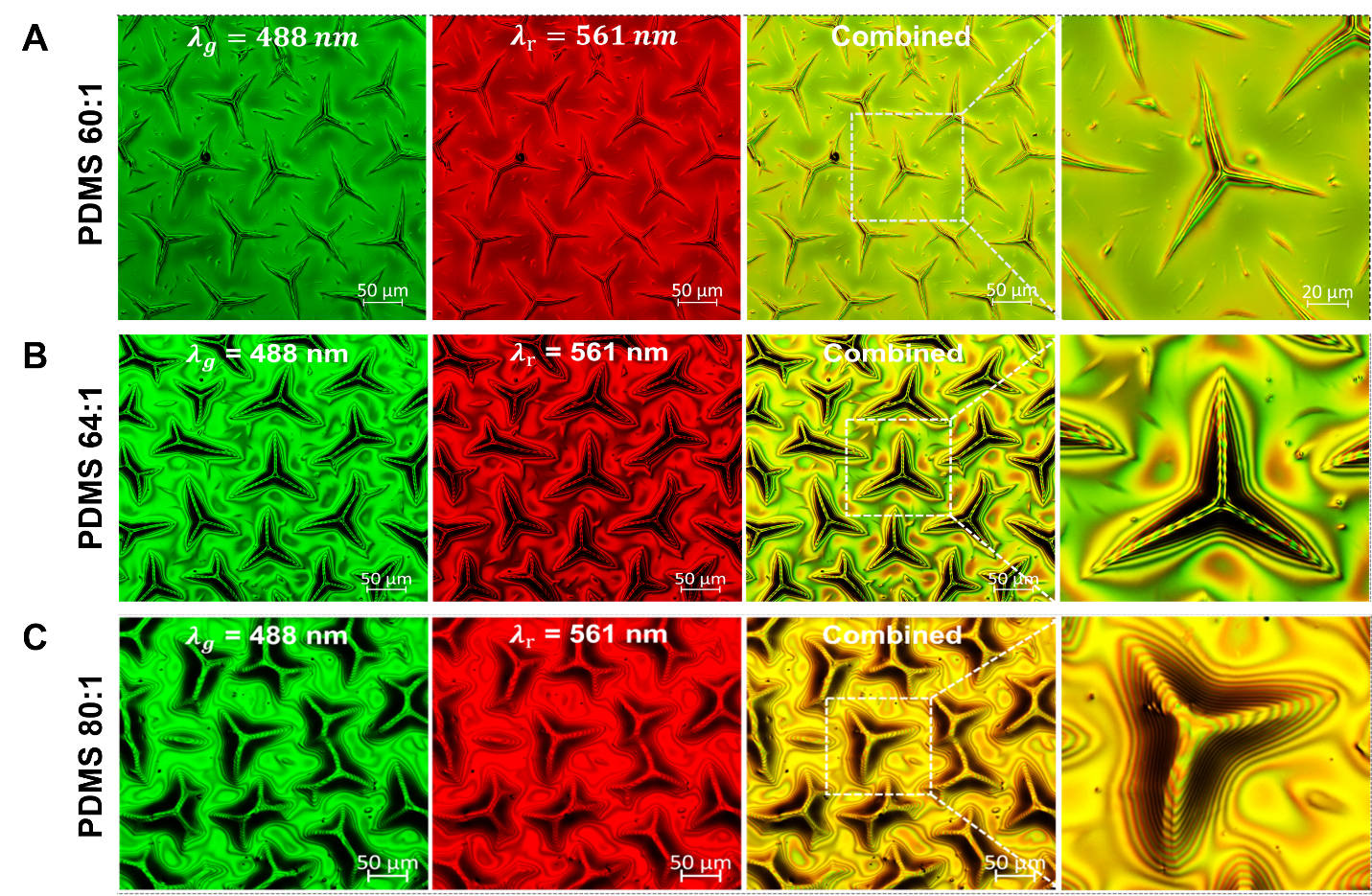}
\caption{Pattern formation for PDMS substrates with different base-crosslinker ratios, swollen in hexane and air-dried: (A) PDMS 60:1, (B) PDMS 64:1, and (C) PDMS 80:1. The green (first column) and the red channels (second column) reflect imaging using \SI{488}{\nano\metre} and \SI{561}{\nano\metre} wavelengths, respectively. The combined channel image (third column) is the overlay of the green and red channel images. The final image in each row is the zoomed-in image of a pattern. Image A adapted with permission from Ref.~\cite{Zhao2025}.}
\label{figS8}
\end{figure}

\begin{figure}[h]
\includegraphics[width=0.85\linewidth]{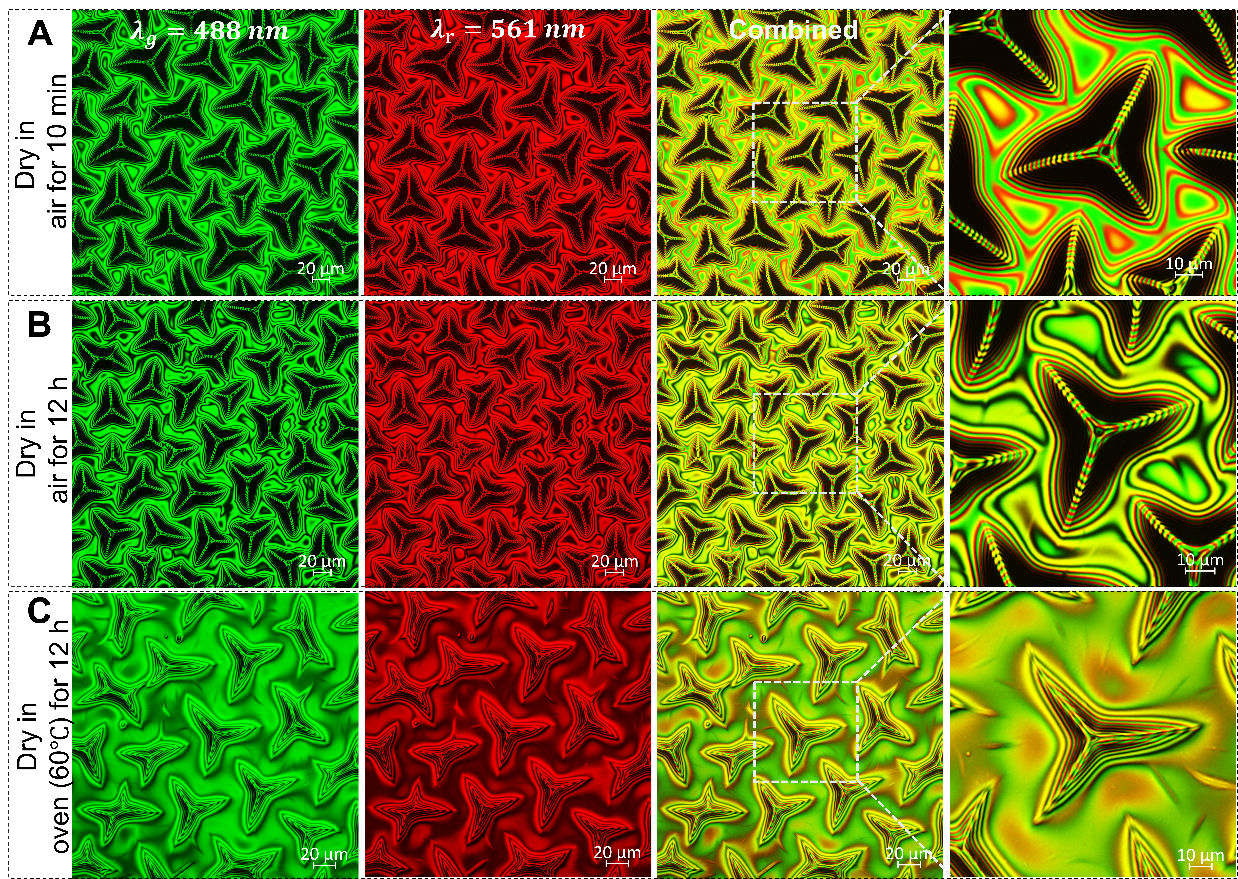}
\caption{Effect of drying conditions on the pattern formation on silicone gel films: (A) in air for 10 mins, (B) in air for 12 hr, and (C) in an oven for 12 hr at \SI{60}{\celsius}. The green (first column) and red channels (second column) reflect imaging using \SI{488}{\nano\metre} and \SI{561}{\nano\metre} wavelengths, respectively. The combined channel image (third column) is the overlay of the green and red channel images. The final column images are zoomed-in versions of the combined channel images, highlighting a single pattern. Image adapted with permission from Ref.~\cite{Zhao2025}.}
\label{figS9}
\end{figure}

\begin{figure}[h]
\includegraphics[width=0.85\linewidth]{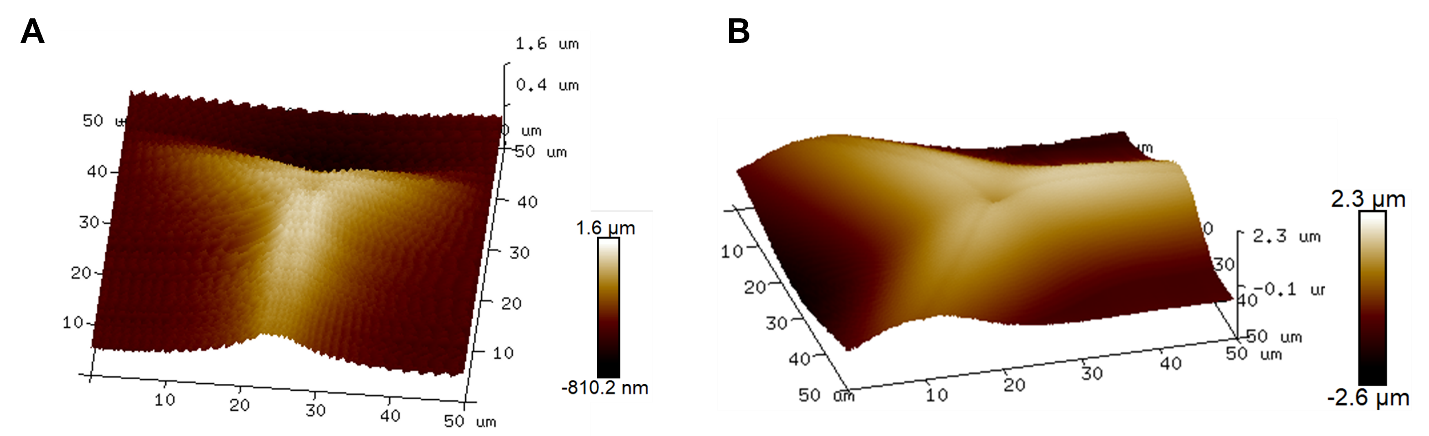}
\caption{(A), (B) Representative AFM scans of single motifs on the silicone gel, from which $h\approx\SI{2.5}{\micro\metre}$ is taken.}
\label{figS10}
\end{figure}

\end{document}